# Reversible quantum optical data storage based on resonant Raman optical field excited spin coherence

**Byoung S. Ham**

*Center for Photon Information Processing, and the Graduate School of Information and Telecommunications,
Inha University, 256 Yonghyun-dong, Nam-gu, Incheon 402-751, S. Korea*
bham@inha.ac.kr

**Abstract:** A method of reversible quantum optical data storage is presented using resonant Raman field excited spin coherence, where the spin coherence is stored in an inhomogeneously broadened spin ensemble. Unlike the photon echo method, present technique uses a $2\pi$ Raman optical rephasing pulse, and multimode (parallel) optical channels can be utilized, where the multimode access gives a great benefit to both all-optical and quantum information processing.

## 1. Introduction

Photon echoes were introduced for multidimensional optical data storage [1,2]. The photon echo technique uses an inhomogeneously broadened two-level optical transition of atoms to store consecutive (quantum) optical information into reversible optical spectral gratings, where the consecutive optical data can be superimposed on the same group of atoms forming multiple atomic spectral gratings. Even though the photon echo method implies potentially ultrahigh-speed all-optical data storage, the photon echo memory has a critical defect of low retrieval efficiency of around 0.1 % and a requirement of extremely low operation temperatures. In 2001, however, a modified photon-echo-based quantum memory utilizing backward retrieval geometry for near perfect retrieval efficiency has been suggested [3]. Even an electro-optic retrieval method using photon echo technique has been experimentally proved for high retrieval efficiency in a rare-earth doped crystal [4-6]. Although the usage of a DC electric field for rephasing of inhomogeneously broadened atoms is simple and has potential for high retrieval efficiency in comparison with the conventional photon-echo technique, the quantum data-bit rate must be limited to the DC voltage confined narrow Stark gradient.

The essence of quantum memory in quantum information processing and communications is found in potential applications to quantum optical devices such as quantum repeaters [7,8], quantum switches [9,10], and quantum logic gates [11]. For these quantum devices, longer quantum storage time, faster data-bit rate with shorter processing time, and/or higher retrieval efficiency are essential components to be considered for practical implementations. Precise manipulations of the temporal shape of the data pulse as well as an auxiliary coupling pulse can enhance the quantum retrieval efficiency up to 50% in the electromagnetically induced transparency (EIT) [12]-based slow-light quantum memory technique [13-16]. However, the quantum data-bit rate and the quantum storage time have been severely limited. In this paper we propose a new quantum optical data storage method using spin inhomogeneous broadening to overcome conventional limitations in retrieval efficiency and data-bit rate.

The physics of the present optical data storage is in a two-photon excited dark state coherence, where the dark state is formed in a reversible spin inhomogeneously broadened ensemble [17-19]. Contrary to previously demonstrated quantum optical methods using optical inhomogeneous broadening such as in photon echoes, the present method utilizes spin inhomogeneous broadening for serial quantum optical data storage as well as optical inhomogeneous broadening for multimode parallel processing (or multiple channel access), where multimode operation gives a great advantage in quantum repeaters. Unlike the photon echo method, the present rephasing mechanism needs a $2\pi$ pulse area [18].

For the rephasing pulse, a resonant Raman optical field is related, where the $2\pi$ rephasing pulse does not have to be equally distributed into both optical fields. This means that there is flexibility of power distribution to the Raman optical rephasing fields. Moreover, absolute optical frequency of each field does not play an important role for the proposed quantum optical data storage and retrieval efficiency because only two-photon phase is related with the spin information. In addition it should be noted that the storage technique is fundamentally different from the off-resonant Raman scattering method, where the off-resonant Raman method relies on a spontaneous decay process [20,21]. The present quantum memory technique does not have to rely on the slow light phenomenon for data storage purpose but benefits from enhanced atomic coherence and enhanced nonlinearity for the retrieval process into optical photons [22,23].

## 2. Theory

In a three-level system a Raman field composed of $\Omega_P$ and $\Omega_C$ induces superposed states $|-\rangle$ and $|+\rangle$, respectively, which are decoupled and coupled from the excited sate $|3\rangle$:

$$|-\rangle = c_1|1\rangle - c_2|2\rangle, \quad (1\text{-a})$$
$$|+\rangle = c_2|1\rangle + c_1|2\rangle, \quad (1\text{-b})$$

where $c_1$ and $c_2$ are $\Omega_2/\Omega$ and $\Omega_1/\Omega$, respectively, and $\Omega^2 = \Omega_1^2 + \Omega_2^2$. The decoupled state is called dark state because no absorption is occurred. The existence of the dark state is the basis of coherent population trapping [24] or EIT [12]. However, such a decoupled state under coherent population trapping, needs a long interaction time $\tau$, in which $\tau \gg 1/\gamma$ ($\gamma$ is an optical decay rate). For EIT, such an interaction time $\tau$ can be shorter ($\tau \ll 1/\gamma$) if $\Omega \gg \gamma$. In this paper we are not interested in none of these cases for the storage process: $\tau < 1/\gamma$; $\Omega < \gamma$. What is considered is full absorption by the medium. As long as optical field absorption is related via a resonant Raman field, spin coherence must be involved. For numerical calculations, we introduce density matrix equations of motion in an interacting Hamiltonian picture [25]:

$$\frac{\partial \rho}{\partial t} = -\frac{i}{\hbar}[H, \rho] + \text{decay terms}, \quad (2)$$

where $H$ is the interaction Hamiltonian. From Eq. (2), a total 9 of density matrix equations of motion is obtained. For the numerical calculations, experimental values of a rare-earth $Pr^{3+}$ doped $Y_2SiO_5$ are used. Here, we present the potential of the present reversible quantum optical data storage [22,23], where multiple optical channels can be utilized in an optical inhomogeneous broadening for an overall data-bit rate of ~Gbps. This is very important for quantum repeater applications, in which multiple quantum memory processes are needed.

## 3. Results and Discussions

Figure 1 shows numerical simulations of the reversible quantum optical data storage using spin inhomogeneous broadening based on resonant Raman optical field excited spin coherence. Figure 1(a) represents a schematic diagram of a typical lambda-type three-level optical system interacting with a pair of resonant Raman fields composed of $\Omega_P$ and $\Omega_C$ at equal strength. All numerical parameters are from experimental values of $Pr^{3+}$ doped $Y_2SiO_5$ at liquid helium temperatures. For simplicity, optical transitions are assumed to be homogeneous. With an inhomogeneously broadened spin ensemble, the resonant Raman fields must experience a two-photon detuning $\Delta_t$ for each atom (spin) group $i$ depending on its spectral detuning. The spin inhomogeneous broadening is Gaussian distributed with 100 kHz of FWHM in state $|2\rangle$ of Fig. 1(a). For the numerical calculations, a total of 121 groups of atoms (spins) divided symmetrically at a 2 kHz step is assumed. The rephrasing pulse R, whose pulse area is $2\pi$ with a 5 ns pulse length, follows the Raman data pulse D whose pulse length is 10 μs as shown in Fig. 1(b). Figures 1(c) and 1(d) represent the Raman pulse excited spin coherence (dark state) for individuals and overall. Figure 1(e) is for the imaginary components of individual atoms. As seen in Fig. 1(c) and 1(e), the phase of the excited atoms runs freely at the end of the Raman data pulse at different speed depending on the detuning $\Delta_t$. Thus, the overall phase quickly dephases as shown in Fig. 1(d).

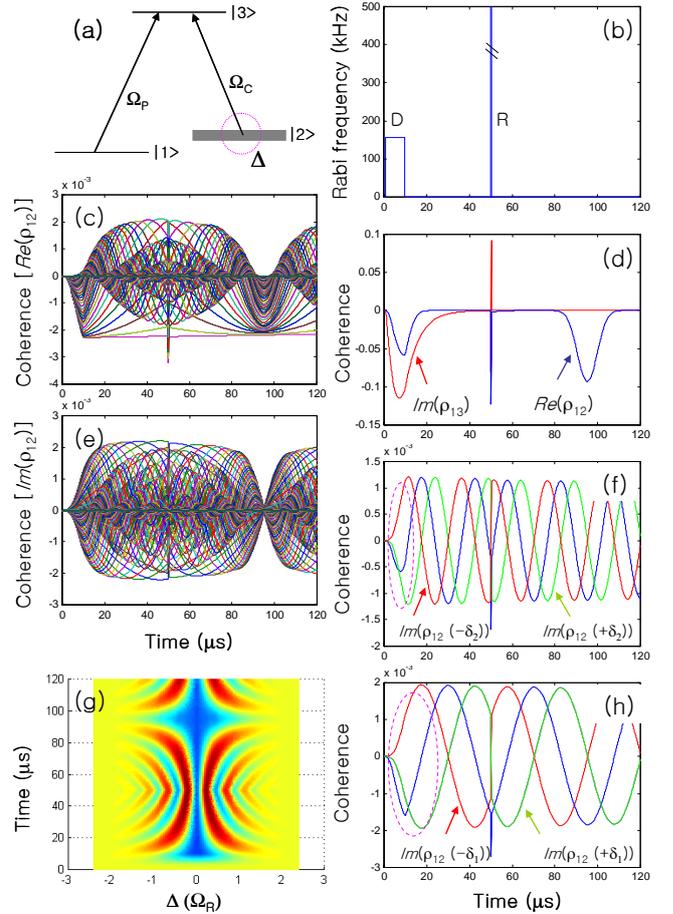

Fig. 1. Resonant Raman optical pulse excited spin coherence and retrieval. (a) Energy level diagram, (b) pulse sequence, (c) spin coherence evolution with a $2\pi$ rephasing pulse, (d) overall spin (blue) and optical (red) coherence, (e) optical coherence evolution, (f) individual spin coherence for symmetrically detuned spins, (g) top-view of three dimensional plot of (c), (h) Spin coherence evolution for closely off detuned spin pairs. $\Omega_P = \Omega_C = 50$ for D; $\Gamma_{31}=\Gamma_{32}=1$; $\gamma_{31}=\gamma_{32}=25$; $\Gamma_{21}=0$; $\gamma_{21}=1$; $\Delta=100$; $\delta_1=4$; $\delta_2=10$ kHz. $\rho_{11}=\rho_{22}=0.5$.

By applying the rephrasing pulse R at t=T, whose function is to make a time reversal in the evolution process, the spin coherence of each atom (spin) becomes retrieved when $t=2T-T_D$, where $T_D$ is the data pulse arrival time. The overall evolution of the coherence retrieval process is shown in Fig. 1(d). When the excited spin coherence is completely retrieved, the absorption component of each spin becomes zero as shown in Fig. 1(e). This proves that the spin coherence can be excited by a resonant Raman optical pulse, and the excited spin coherence can be retrieved by a $2\pi$ optical rephasing pulse. Figure 1(g) shows top view of three-dimensional picture of the coherence excitation and retrieval: In the rainbow color, the red (blue) stands for positive (negative) values. Because of the time reversed process by the action of rephasing pulse at t=50 µs, the initial spin coherence is retrieved at t=95 µs as discussed in Figs. 1(c) and 1(d). For a particular group of atoms (spins) more detuning results in a faster oscillation in the evolution process. In Figs. 1(f) and 1(h), the real component of the spins shows symmetry across the rephasing pulse, while the imaginary component shows antisymmetry. Like in the response function, the phase of imaginary parts of the spins either precedes or lags by $\pi/4$ against the phase of the real component. Because of this relationship between the real and imaginary terms, the magnitude of the very first (front surface) excited spin coherence for the real term is always less than the others (see the dashed circles). This is why the retrieved coherence has greater value than the excited one in Fig. 1(d), but cannot be greater than the absorption value of the optical coherence $Im(\rho_{13})$ to satisfy the energy conservation rule.

Figure 2 shows coherence evolutions of all spins interacting with the resonant Raman pulses of data D and rephasing pulse R. Figure 2(a) is an extended feature of Fig. 1(c). Figure 2(b) is for all spins in the uv plane of real and imaginary components of the spin coherence at t = 10 µs, where u (v) represents real (imaginary) part of the spin coherence $\rho_{12}$. Figure 2(c) is an extended feature of Fig. 1(c), where all spins are dephased completely. As shown in Fig. 2(d), all spin coherence components becomes random at t=45 µs (see also Fig. 1(d)). Figure 2(e) is an extended feature of Fig. 1(c) and Fig. 2(f) is in the uv plane, in which all spin phases are retrieved at t = 90 µs. In comparison with Fig. 2(b), Fig. 2(f) shows almost same pattern demonstrating coherence retrieval as shown in Fig. 1(d). The maximum coherence retrieval occurs at 95 µs because the data pulse length is 10 µs.

Figures 3 shows numerical simulations of the present quantum memory based on resonant Raman field excited dark state coherence using reversible spin inhomogeneous broadening. The inset of Fig. 3(a) shows a schematic of a typical lambda-type three-level optical system interacting with a pair of the resonant Raman fields composed of $\Omega_P$ and $\Omega_C$ at equal strength. Figure 3(a) represents the resonant Raman pulse sequence: the first two pulses stand for data (D1 and D2) and the third pulse stands for rephasing (R). For simplicity, optical transitions are assumed to be homogeneous, and the value of the optical homogeneous decay rate is effectively affected by the laser jitter [26]. With an assumption of an inhomogeneously broadened spin ensemble, the resonant Raman fields give a two-photon detuning $\Delta$ to the inhomogeneously broadened spin ensemble. The spin inhomogeneous broadening allowed to the state |2> is assumed to be Gaussian distributed at 100 kHz (FWHM). For the calculations, a total of 121 groups of atoms (spins) in the inhomogeneous broadening are equally divided at a 2 kHz step. The rephasing pulse R whose pulse area is $2\pi$ with a 5 ns pulse length follows the Raman data pulse D1 and D2 whose pulse length is 10 µs as shown in Fig. 1(a).

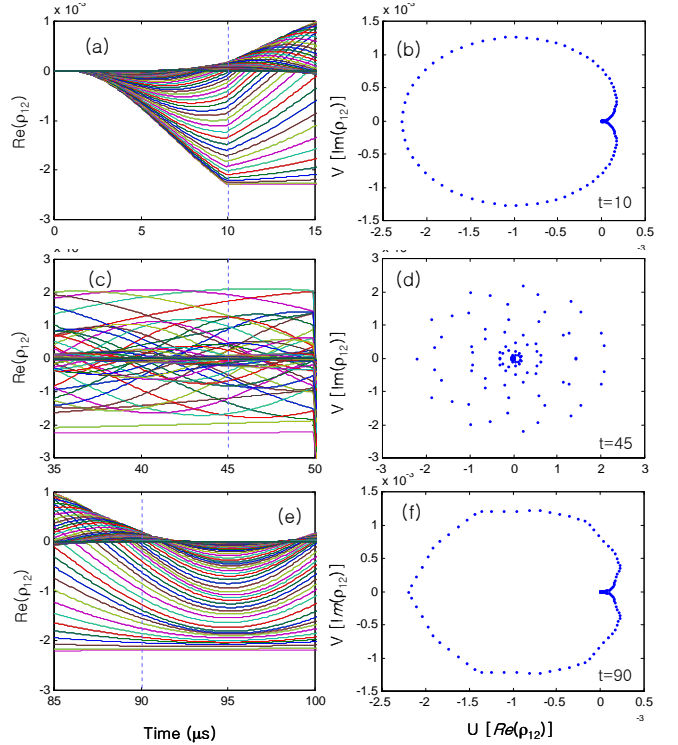

Fig. 2. Coherence rephasing for quantum optical data storage. (a, c, e) extended feature of Fig. 1(c), (b, d, f) bloch vector diagram in a uv plane for spin coherence excitation, dephasing, and rephasing.

Figure 3(b) represents the present quantum memory, where the consecutive Raman data pulses (D1 and D2 in Fig. 1(a)) excite corresponding spin coherences (dark states, DC1 and DC2), and the Raman rephasing pulse retrieves the excited spin coherences back (DR1 and DR2) in a time-reversed manner. At the end of each data pulse, the excited spin coherence dephases much faster than the spin phase decay time $T^S_2=1/(\pi\gamma^S_2)=318$ µs. This is the well-known phenomenon of free induction decay, where the dephasing time is inverse to the spin inhomogeneous broadening. For the rephasing pulse area of $2\pi$ (blue curve) or $6\pi$ (green curve), the retrieval efficiencies are the same. However, the $4\pi$ rephasing pulse (red curve) can never retrieve the stored coherence back (will be discussed in Fig. 3(c)). The inset of Fig. 3(b) shows an individual spin coherence evolution whose detuning δ is ±10 kHz to the $\Omega_C$.

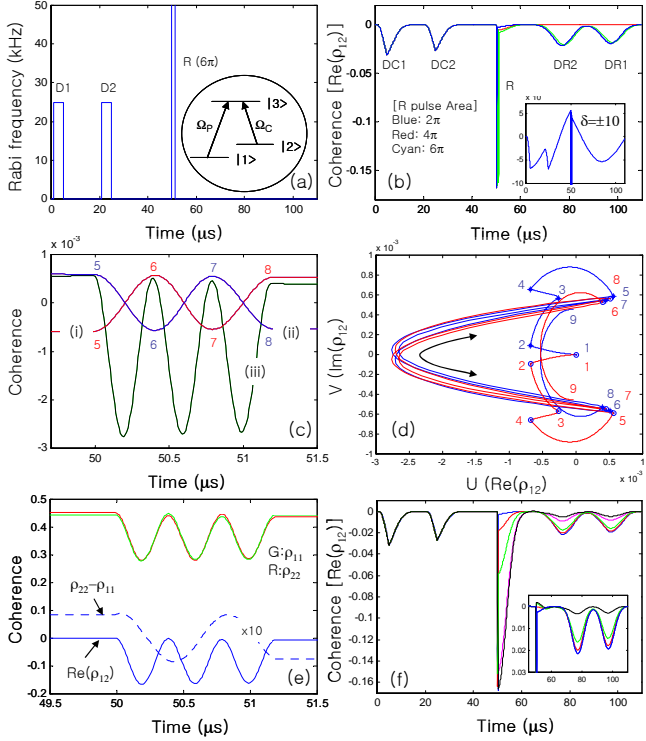

Fig. 3. Multiple quantum optical data storage. (a) Raman optical pulse sequence, (b) consecutive quantum optical data storage, (c, d) Rephasing mechanism, (e) population vs coherence, (f) rephasing pulse area dependence. Inset: rephasing pulse ratio of $\Omega_P$ to $\Omega_C$, blue: 50/50, red:30/40, green: 25/43.3, black:10/49

Figure 3(c) shows the rephasing process at around t=50 μs for the inset of Fig. 3(b). The red curve (i) and the blue curve (ii) represent the imaginary parts of the spin coherence for symmetric detuning with δ=10 kHz and δ=−10 kHz, respectively. The black curve (iii) represents the real part of both spin coherences whose detuning is δ = ±10 kHz. For the first 2π rephasing pulse area (from "5" to "6") the phase of the real part is retrieved, while the imaginary parts are reversed. Here it should be noted that the reversed phase undergoes a time-reversal process for a complete coherence retrieval as in the photon echo theory. For the 4π rephasing pulse area (from "5" to "7"), however, both the real and imaginary parts recover their coherences completely. This means that the 4π rephasing pulse does not change anything in the spin coherence evolution, so that a complete dephasing results. Because the imaginary part plays a main role for the rephasing process, and the oscillation period of the imaginary component is half of the real part, an additional 4π pulse area is needed for the same result. Thus, a 6π rephasing pulse gives the same retrieval efficiency as the 2π rephasing pulse does. This means that the minimum rephaisng pulse area is 2π and can be increased on a 4π basis.

Figure 3(d) shows the resonant Raman pulse excited spin coherence in a uv plane of the Block diagram for the inset of Fig. 3(b). By the action of the first Raman data pulse D1, the spins whose detuning is δ = ±10 kHz are gradually excited ("1"-"2") and then diphase ("2"-"3") until the second Raman data pulse D2 comes. Here the red (blue) represents for positive (negative) detuning. For the second Raman data pulse D2, the same spins re-excite gradually ("3"-"4") and then dephase again ("4"-"5"). When the rephasing pulse R (6π) is applied, the spin coherence experiences oscillation depending on the rephasing pulse area, where the real (imaginary) part has 2π (4π) modulo as shown in Fig. 3(c). This means that the uv plane is flipped 180 degrees along the u axis with a 2π rephasing pulse area basis ("5"-"6", "6"-"7", and "7"-"8"). Thus, like the photon echo technique, the coherence becomes retrieved at t=2T, where T is the temporal elapse between the data pulse and the rephasing pulse R. Because the rephasing pulse turns each spin coherence backward in the time domain, the retrieved signals must be timely reversed (shown in Fig. 2). Unlike the photon echoes, the present method shows that the retrieved coherence keeps the same polarization as the excitation polarization (see Fig. 5). This is because photon echo relates with imaginary part of coherence (will be discussed below and in Fig. 5).

Now we discuss the rephrasing process in more detail. In Fig. 3(e), the spin coherence evolution of Fig. 3(b) is compared with the populations on both ground states |1> and |2>. Each state population ($\rho_{11}$ or $\rho_{22}$) oscillates according to the spin coherence (blue curve). The population difference ($\rho_{22}-\rho_{11}$) between states |2> and |1> behaves like the imaginary part of the spin coherence (dotted curve), resulting in a complete population reverse (see the sign change for the first 2π pulse area). This is the same effect that a resonant rf π-pulse applies to the transition |1> − |2> resulting in an upside-down population. Thus, we conclude that the 2π Raman optical pulse plays a role in a π rf pulse, causing the coherence polarization reversed as shown in Fig. 3(c) and 3(d).

Figure 3(f) shows the flexibility of using the rephasing Raman optical pulse. The blue curve shows the 2π pulse area. The red and green curves represent 10% and 20% pulse area reduction, respectively. The corresponding retrieval efficiencies are 91% and 77% of the 2π pulse area case, respectively. Even for the cases of 40% (magenta curve) and 50% (black curve) pulse area reduction, retrieval efficiency reaches up to 41% and 27%, respectively. The inset of Fig. 3(f) shows retrieval efficiencies for different ratios of $\Omega_P$ to $\Omega_C$ of the 2π rephasing Raman pulse R. As shown the retrieval efficiency is sensitive to both pulse area and the field ratio of the Raman pulse.

Figure 4 shows coherence transfer for quantum optical memory, where the atomic (spin) coherence excited by a resonant Raman optical pulse in Fig. 1(d) is extracted as a photon emission via coherence transfer processes [27]. Figure 4(a) shows individual coherence evolutions of the spin ensemble by adding a readout pulse of $\Omega_C$ in addition to Fig. 1(c). Figure 4(b) shows photon retrieval of $\Omega_P$ by the readout pulse $\Omega_C$. The atomic coherence Re$\rho_{12}$ is converted into photons of $\Omega_P$ via nondegenerate four-wave mixing processes [17-19,22,23,27]. Figure 4(c) is the sum of Figs. 4(a) and 4(b): the red (blue) curve is for photon emission (atomic coherence). In Fig. 4(d) the atomic coherence excitation, rephasing process, and coherence conversion process are presented in comparison with population change. At t=40 µs, the population between states |1> and |2> seems to remain unchanged with the resonant $2\pi$ Raman optical rephasing pulse. However each state population is actually swapped via the excited state |3>, as a $\pi$ pulse is in a two-level system does if two state are equally populated (discussed in more detail in Fig. 5). The action of the readout pulse at t=70 µs is to depopulate the state |2>. Here, the readout pulse transfers the population on state |2> to the state |1> via |3>. To bring about efficient photon emission, the medium must be nonabsorbing. The nonabsorbing medium to the resonant optical field can be obtained by EIT as demonstrated [22,23].

atomic coherence $\rho_{12}$. In the uv plane any freely decaying spin should rotate in a clockwise or counterclockwise direction, depending on its sign of two-photon detuning (see Fig. 3(d)). For a positive detuning $\delta\iota$, the spin rotates in a clockwise direction at a fixed speed. For coherence excitation by the resonant Raman optical pulse (a−b), the real component of $\rho_{12}$ gradually increases, as shown in Fig. 4(a). For the rephasing by the $2\pi$ resonant Raman optical pulse (dotted region), only the phase of $\rho_{12}$ is shifted by $\pi$ as discussed in Figs. 3(c) and 3(d). This process makes the atomic coherence reversed as discussed in Figs. 3. When the time reaches point "p," where the phase is exactly opposite of point "b," the readout pulse is applied to deplete the spin coherence. The coherence of Re$\rho_{12}$ becomes zero during the readout pulse in the p–q region. Figure 4(f) shows a top view of three-dimensional plot of photon absorption and emission in Fig. 4(a). As discussed the coherence conversion by the read-out pulse $\Omega_C$ induces photon emission (red color).

Figure 5 shows the rephasing process in comparison with the photon echo [2]. Figure 5(a) shows individual atomic coherence evolution. Two consecutive Raman optical data pulses are for the regions of "a−b" and "c−d." The $2\pi$ rephasing pulse is applied at t=50 µs for 400 ns to satisfy the $2\pi$ pulse area: "e-f-g." During the rephasing process, however, the population difference between states |1> and |2> is swapped completely, causing

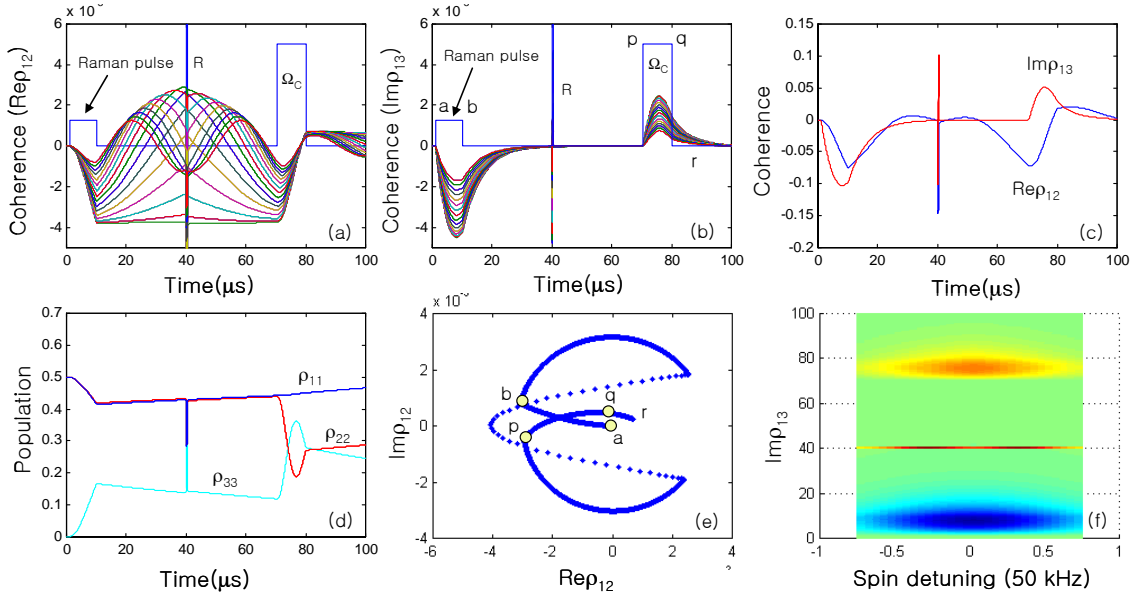

Fig. 4. Coherence conversion between photons and spins. (a) spin coherence depletion by the action of readout optical pulse, (b) optical emission generation by the action of readout pulse, (c) overall coherence transfer between photons and spins, (d) population evolution, (e) single spin coherence evolution, (f) top-view of three dimensional plot of (b).

Figure 4(e) shows a particular spin evolution in the uv plane composed of real and imaginary parts of spin phase reversed (see Figs. 5(b) and 5(c)): The small difference is due to equal population distribution with inhomogeneous broadening. If the population distribution ratio is big, the effect must be big enough (see Fig. 5(f)).

Figure 5(d) shows typical photon echo simulations for both the $\pi$ (red curve) and $2\pi$ (blue curve) rephasing pulse area. For this the optical phase decay rate is adjusted, otherwise no rephasing is possible: 25 kHz → 2.5 kHz. For the $2\pi$ rephasing pulse, the echo disappears completely due to the complete phase recovery resulting in zero phase shift. Figure 5(e) shows a top view of three-dimensional plot of the coherence evolution of Fig. 5(d) with a $\pi$ rephasing pulse. Thus, it is clearly that a $2\pi$ resonant Raman optical pulse induces the spin rephasing process as a $\pi$ pulse does in the photon echoes.

is set much wider than both $\Omega$ and $\gamma_{OPT}$, the effect from the neighboring group of atoms can be neglected. In this case the number of optical channel modes N is determined by $N \sim \frac{\Delta_{OPT}}{\Omega}$. For $Pr^{3+}$ doped $Y_2SiO_5$, N~1000. This means that 1000 parallel entanglement swapping operations in a quantum repeater can be possible [28].

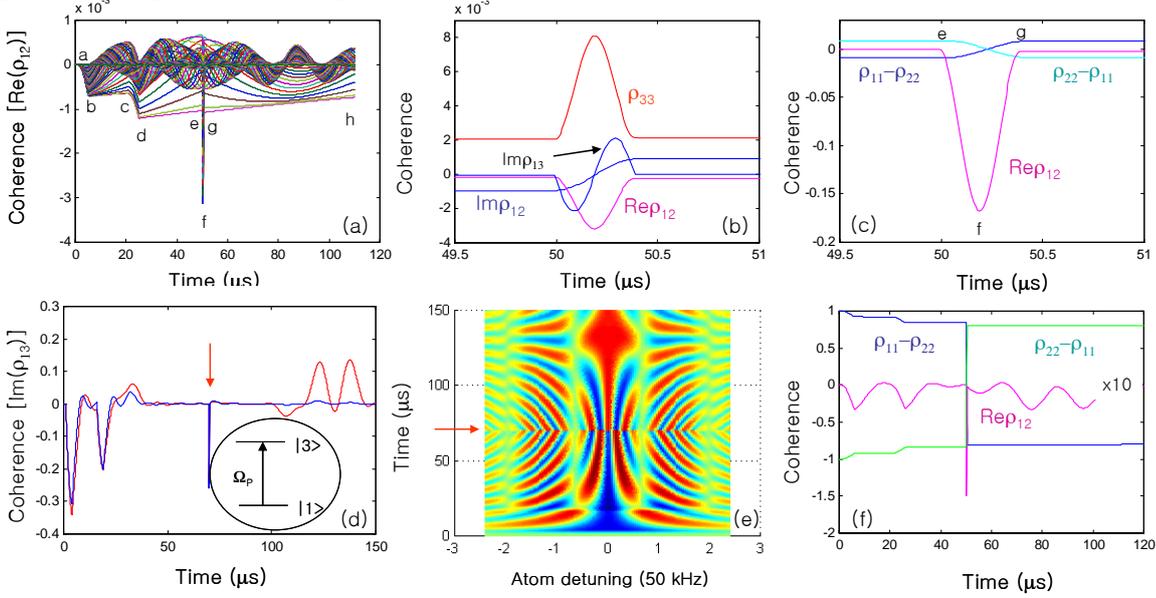

Fig. 5. Rephasing mechanism. (a) individual spin coherence evolution of Fig. 3(a), (b) coherence change by the action of $2\pi$ rephasing pulse, (c) population difference versus spin coherence rephasing, (d) photon echo with $\pi$ (red) and $2\pi$ (blue) rephasing pulse, (e) top-view of individual atoms of (d) with $\pi$ rephsing pulse, (f) spin coherence versus population difference for $\rho_{11}=1$; $\rho_{22}=0$.

Now we show two-pulse quantum optical memory. Everything is same as in Fig. 5 except the readout pulses A1 and A2 in Fig. 6(a). As shown in Fig. 6(b), the photon emission ($Im\rho_{13}$) is retrieved from the atomic coherence conversion process as discussed in Fig. 4. Figures 6(c) and 6(d) represent three-dimensional plots of atomic and optical coherence evolutions. Unlike Fig. 4(f), the photon emissions in Fig. 6(d) show spectral modulations. The spectral modulation results from the consecutive Raman optical data pulse sequence via the rephasing pulse (Fourier transformation). Here it should be noted that the readout sequence may be backward, because the readout pulse may influence the spin coherence chain [27].

For multimode (parallel) quantum optical data storage, optical inhomogeneous broadening $\Delta_{OPT}$ is utilized (see Fig. 7). Each optical mode (channel) spectral width can be adjusted according to the relation among optical field Rabi frequency $\Omega$ and optical phase decay rate $\gamma_{OPT}$. If the spectral spacing between neighboring groups of atoms

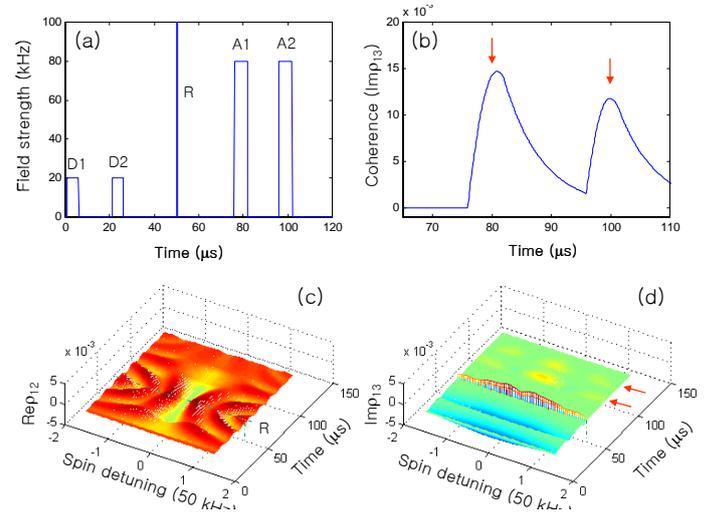

Fig. 6. Multiple quantum optical data storage and retrieval. (a) pulse sequence, (b) consecutive coherence transfer from spin coherence to optical emission, (c) three-dimensional plot of spin coherence, (d) three-dimensional plot of optical coherence.

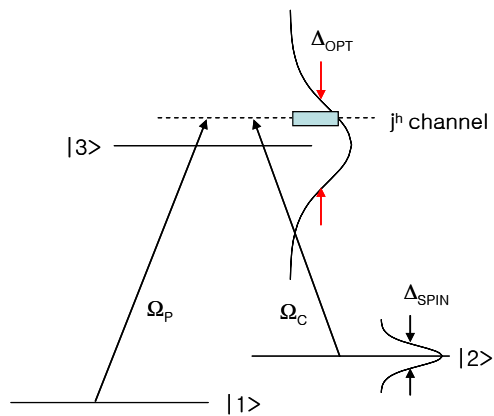

Fig. 7. Schematic diagram of multimode quantum optical data storage.

## 4. Conclusion

Resonant Raman pulse excited atomic coherence and rephasing processes are investigated for quantum optical data storage, where the retrieval process is time-reversed and multimode optical pulses are used. Unlike the photon echo technique, the present method holds a $2\pi$ Raman rephasing pulse area. The multimode (parallel) quantum optical data storage can also be obtained using optical inhomogeneous broadening, where the multimode quantum memory is advantageous in both classical and quantum optical memories.

## Acknowledgments

This work was supported by Creative Research Initiative Program (Center for Photon Information Processing) of MEST via KOSEF, S. Korea.